\documentclass[prl,aps,floatfix,twocolumn,showpacs,preprintnumbers,superscriptaddress]{revtex4}
\usepackage{graphicx}
\usepackage{dcolumn}
\usepackage{bm}
\usepackage{amssymb}
\usepackage{pifont}
\usepackage{amsmath}
\usepackage{times}
\usepackage[nooneline]{subfigure}
\begin{document}

\title{Optimization and plasticity in disordered media}

\author{Clara B.\ Picallo} \email{picallo@ifca.unican.es}
\affiliation{Instituto de F\'{\i}sica de Cantabria (IFCA), CSIC--UC, E-39005
Santander, Spain}
\affiliation{Departamento de F{\'\i}sica Moderna, Universidad de Cantabria,
Avda. Los Castros, E-39005 Santander, Spain}
\author{Juan M.\ L{\'o}pez} \email{lopez@ifca.unican.es}
\affiliation{Instituto de F\'{\i}sica de Cantabria (IFCA), CSIC--UC, E-39005
Santander, Spain}

\author{Stefano Zapperi} \email{stefano.zapperi@unimore.it}
\affiliation{CNR-INFM, S3, Dipartimento di Fisica, Universit{\`a} di Modena e
Reggio Emilia, 41100 Modena, Italy}
\affiliation{ISI Foundation, Viale S. Severo 65, 10133 Torino, Italy}

\author{Mikko J.\ Alava} \email{mikko.alava@tkk.fi}
\affiliation{Department of Applied Physics, Helsinki University of Technology, 
HUT-02015, Finland}

\date{\today}

\begin{abstract}
We study the plastic yielding of disordered media using the perfectly plastic
random fuse model. The yield surfaces are shown to be different from those
obtained minimizing the sum of the local yield thresholds, {\it i.e.} the
so-called
minimum 'energy' surfaces. As a result, the global yield stress is lower than
expected from naive optimization and the difference persists as the sample size
increases. At variance with minimum energy surfaces, height-height fluctuations
of yield surfaces exhibit multiscaling. We provide a theoretical
argument that explains how this behavior arises from the very different nature
of the optimization problem in both cases.
\end{abstract}


\maketitle

When subject to large loads, materials can deform plastically, changing
irreversibly their shape. Macroscopically, this process is described by the
continuum theory of plasticity, stating that at the yield stress the sample
develops plastic strain. There has been much interest in the so-called perfect
plasticity (PP) limit, when plastic strain grows without any further increase of
the external stress. In crystalline materials yielding is explained as the
motion of dislocations in response to the applied stress~\cite{zaiser}. In
contrast, yielding   in amorphous materials is due to irreversible atomic
rearrangements. The latter has been mostly studied by means of extensive
molecular dynamics simulations~\cite{falk98,falk99,zink06,lem07}. The insight
gained through the numerics has led to mean field descriptions based on
localized events in shear transformation zones (STZ) (see~\cite{falk98,lang08}
and references therein).

Bridging gap between the length scales of microscopic models and continuum
theories is yet one of the most challenging problems in materials science. The
main difficulty for homogenization processes stems from the strong localization
of plastic strain into slip lines ---in crystals ---or shear bands--- in
amorphous media. Nevertheless, some efforts have been made to study plastic
deformations at mesoscopic scales~\cite{sor93a,roux02,picard05,pp92}. In this
framework, the yield surface results from the joint optimization of local
intrinsic disorder and elasticity. The presence of local stress thresholds has
been shown to induce the appearance of localization into shear bands.

Based on a powerful analogy, it is generally believed~\cite{pp91,pp92} that
strain localization in the PP limit can
be related to the problem of finding the minimum
energy (ME) surface in a disordered medium.
This is a generic optimization problem in disordered media in which
one searches for the path that minimizes the sum of a given local random
variable that is called 'energy'.
The conjectured equivalence between
PP and ME comes from the observation that, at the yield point, it is not
possible to find an elastic path, along which the stress could increase, spanning the
sample from end to end~\cite{pp91,pp92}. In a disordered medium, the local yield
stress $\sigma_i$ of a given cross-section
is in general a quenched random quantity. Therefore, according to
Refs.~\cite{pp91,pp92}, the global yield stress $\sigma_c$
could be obtained by finding the surface $\mathcal{S}$ where the sum of the
local yield stresses (the 'energy')
is minimized (i.e. $\sigma_c=\min_{\mathcal{S}}[\sum_{i \in \mathcal{S}}
\sigma_i$]). When this
value of the stress is reached, the system would be divided into two disconnected elastic
parts and would thus behave as perfectly plastic.

ME surfaces in
disordered media have been intensively studied in
the last twenty years since they appear in many contexts and
several results are known exactly~\cite{revDP}. In particular, the ME surface in
two dimensions is equivalent to a directed polymer at zero temperature and is
thus a self-affine object with a roughness exponent $\zeta=2/3$
and an energy exponent $\theta=1/3$, describing
the system size scaling of the energy fluctuations.
The latter implies that yield stress for PP is expected to
display finite size corrections of the type
$\sigma_c = \sigma_\infty +A L^{\theta-1}$.
These corrections are particularly
intriguing since they naturally connect to size effects that have recently been
reported at micron scales both in crystals~\cite{uchdim} and amorphous
materials~\cite{lee07}. In particular, the case of metallic glasses is currently
a fertile ground for research~\cite{lee07,shan08,guo08,schu08,munilla09}.
Microscopic and
nanoscopic samples are known to display way bigger yield strengths and stresses
than bulk samples from the same material but this size-dependence vanishes with
sample diameters only tens of microns larger. This behavior surprisingly fits
the kind of size scaling that ME interfaces display. Nonetheless, recent results
using a shear plane yield criterion revealed size-independent properties at
micron scales as well~\cite{schu08,dubach09}.

In this Letter we argue that the relation between PP and ME should be revised.
By numerical simulations and theoretical arguments we show that PP and ME
actually correspond to two different optimization problems in disordered media.
As a consequence, the yield stress for PP is indeed smaller than the one
observed for the equivalent ME problem, while the critical exponents
of the surface and energy
fluctuations appear to be the same. In two dimensions the yield surfaces have a
roughness exponent of approximately $\zeta=2/3$, and the yield stress
fluctuations scale with an exponent close to the $\theta =1/3$
that corresponds to
the ME universality class. However, the specific surfaces
are different in the two cases. Indeed, the geometry
of the surface in the PP
problem shows the presence of overhangs and large steps that lead to {\it
multiscaling}--- a dependence of the ($q$-th order) roughness exponent on the
order of the correlation function. The presence of overhangs has a significant
effect on the global yield stress. Contrary to what happens in the common ME
problems, overhangs lower the global yield stress so that a trivial minimization
of the sum of local yield stresses is not accomplished.

In our numerical simulations we used the random fuse model
(RFM)~\cite{arcangelis85}, which represents a scalar lattice electrical analog
of the elasticity problem where the stress ($\sigma$), local elastic modulus
($E$), and strain ($\epsilon$) are mapped to the current density ($J$), local
conductance ($g$), and local potential drop ($v$), respectively. The usual
procedure, widely applied to investigate quasi-brittle materials, consists of
fixing the fuse conductivities to unity,  $g_{i}=1$, and assigning to each fuse
a random quenched threshold current $T_{i}$ extracted from e.g. a uniform
distribution~\cite{advphys}.
An external voltage (``strain'') is imposed between two bus bars placed at the
top and the bottom of the system, and periodic boundary conditions are imposed
in the horizontal direction. In studies of brittle fracture the fuses behave
linearly until they fail irreversibly when the local current reaches its
threshold $|J_i| \geq T_i$. However, we are interested here in the plastic
response and thus the local current (local stress) remains constant and equal to
the threshold $|J_i|=T_i$, regardless of the local voltage (strain).

The simulation of the plastic process consists of yield
iterations. At each update, the Kirchhoff equations are solved to
determine the local currents flowing in the lattice. We then increase the voltage up
to the point where the most suscetible fuse yields.
After each yield event, the new currents are computed using the tangent
algorithm
introduced by Hansen and Roux \cite{pp92} and the process is iterated. 
After a large number of iterations, a yield
surface is eventually formed across the
sample. This is the PP yield surface in the sample, which is
univocally determined for each disorder realization.
On the other hand, the corresponding ME surface for the same
disorder
realization is calculated by using the
Edmonds-Karp algorithm~\cite{mid95}.
Note that, contrary to the
brittle RFM, in the PP problem there are no avalanches. This is due to
the fact that there is no current (stress)
enhancement after yield events and thus stresses are not redistributed unlike in
the brittle RFM and other models with avalanches~\cite{advphys}.

For the RFM the need to solve a large system of linear equations for each update
implies a high computational cost and limits the system size and the statistical
sampling. While in the past the best performance was achieved by conjugate
gradient methods~\cite{fourier}, recently, a new
algorithm~\cite{algorithm1,algorithm2} based on rank-1 downdate of sparse
Cholesky factorizations has been introduced, which can largely reduce the
computational cost of the simulations in the RFM. This has allowed to reach
larger system sizes and improve sampling in smaller systems. Here we make use of
this algorithm to study two-dimensional networks of fuses in diamond lattices.
We study systems of linear size ranging from $L=50$ to $L=200$ and $10^{4}$
realizations of the disorder.
\begin{figure}
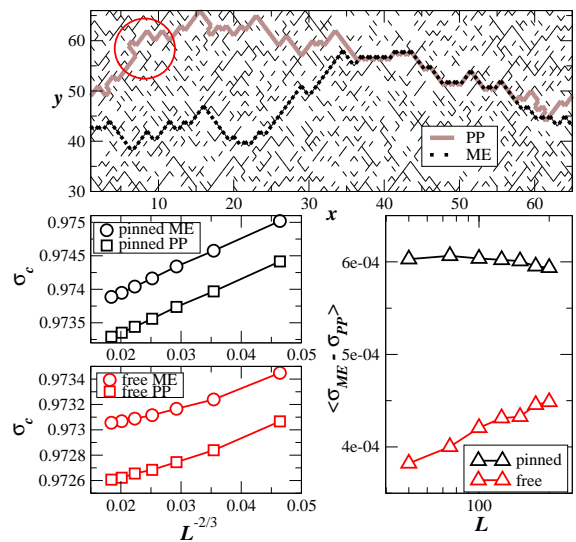

\centerline{\includegraphics *[width=75mm,type=eps,ext=.eps,read=.eps]{fig1}}
\caption{(Color online) Top: A typical ME and PP yield surface for the same
disorder realization in a $L$ = 64 diamond lattice. Bottom: On left panel,
scaling of critical stress with system size in ME and PP for both, fixed and
free ends. Right panel shows that the difference between the critical stress for
ME and PP grows slowly but systematically or remains approximatively constant
with system size for free or pinned ends, respectively.}
\label{fig1}
\end{figure}

Figure~\ref{fig1} (top panel) shows typical ME and PP yield
surfaces for the same
disorder configuration in a typical realization of the RFM. One can clearly see
that the
resulting interfaces may partially overlap but are clearly different. 
In particular, the PP surface presents very visible overhangs. 
As a consequence, the 'energy' of
the PP surface (which corresponds to the sum of thresholds over the yield path)
is indeed higher than that for the ME surface. However, the actual current
(yield stress) through the PP surface is lower than its energy, and also lower
than that for the ME surface.

We claim that the difference between PP and ME surfaces for the {\it same}
disorder realization can be explained by the following theoretical argument. The
equivalent
yield stress for the ME problem in a system of lateral size $L$ is given by 
\begin{equation}
\label{sig_me}
\sigma_{c,ME}= \sum_{i \in \mathcal{S}} T_{i}/L,
\end{equation}
where $i$ runs over all the bonds in the yield surface $\mathcal{S}$ that
minimizes~(\ref{sig_me}). 
In contrast, 
the PP surface $\mathcal{S}'$ would be the surface that requires a lowest
external stress to
appear and, therefore, the one
that minimizes 
\begin{equation}
\label{sig_pp}
 \sigma_{c,PP}= \sum_{i \in \mathcal{S}'} (\mathbf{n}_i \cdot \mathbf{j}_i) \,
T_{i}/L, 
\end{equation}
where $\mathbf{n}_i$ is the unit vector locally normal to the surface at $i$,
and $\mathbf{j}_i=\mathbf{J}_i/|J_i|$ is the local current flow direction.
Eq.~(\ref{sig_pp}) corresponds to the definition of the current flowing through
an
arbitrary surface.
 
If the surface had no overhangs we would have 
$\mathbf{n}_i \cdot \mathbf{j}_i=1$ for all $i$ and the
same surface $\mathcal{S} = \mathcal{S}'$ would minimize both 
Eq.~(\ref{sig_me})
and Eq.~(\ref{sig_pp}). However, in the presence of overhangs, it could happen
that locally $\mathbf{n}_i
\cdot \mathbf{j}_i=-1$ so that the surfaces $\mathcal{S}$ and $\mathcal{S}'$ 
are no longer the same.
Indeed, we find that
$\sigma_{c,PP} < \sigma_{c,ME}$, although the sum of thresholds along the PP
path
is naturally higher than $\sigma_{c,ME}$. 
Therefore, the 
mapping  between minimum energy and yield stress exists only for fully directed surfaces
($\mathbf{n}_i \cdot \mathbf{j}_i=1$ for all $i$), where 
the total yield stress can be calculated as the sum of local yield stresses.
Physically, this means that PP and ME actually
correspond to two different optimization problems.
A PP path may find it very advantageous to develop overhangs in order to
minimize Eq.~(\ref{sig_pp}) due to the negative contributions coming from the
$\mathbf{n}_i \cdot \mathbf{j}_i<0$ terms.
On the contrary, for the ME surface one has to
minimize (\ref{sig_me}) and
overhangs generally increase the global energy and are thus normally avoided, unless
disorder has a very broad distribution~\cite{hav06}.

The difference between the ME and PP yield stresses is quantified in
Fig.~\ref{fig1}. Two different boundary conditions have been studied: the two
ends of the path are either left free or pinned at mid-system. These two
situations correspond to finding either a global or local minimal surface,
respectively. Left panel shows the yield stress scaling with system size for
both free and fixed boundary conditions. In both cases the existence of a finite
size correction becomes apparent, as well as the fact that $\sigma_{c,PP}(L)<
\sigma_{c,ME}(L)$ is always satisfied. For fixed boundary conditions we find
$\sigma_c=\sigma_\infty+AL^{-2/3}$ leading to $\theta=1/3$, which is the
expected result for the ME universality class and likewise so for the PP
problem. Right panel shows the average yield strength difference
$\langle\sigma_{c,ME}-\sigma_{c,PP}\rangle$ that systematically increases with
$L$ for free boundary conditions or remains constant in the case of fixed
boundary conditions. 

\begin{figure}
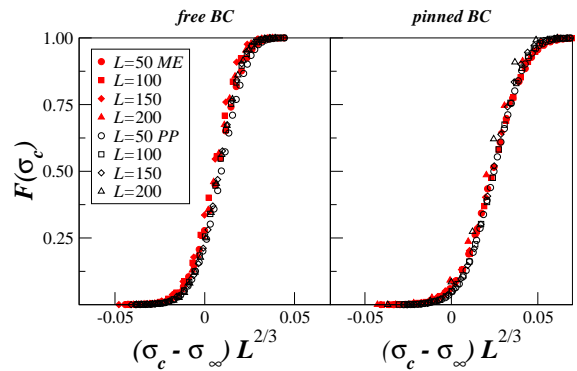

\centerline{\includegraphics *[width=75mm,type=eps,ext=.eps,read=.eps]{fig2}} 
\caption{(Color online) Cumulative distributions of yield stress for free and
pinned boundary
conditions for ME and PP. The distributions can all be collapsed with the same
exponent, related to $\theta=1/3$.} 
\label{fig2}
\end{figure}

The scaling of the yield stress is
reminiscent of size effects, traditionally studied in brittle fracture
problems, where one expects extreme value statistics to
apply~\cite{weibull,jphysd}. Although size effects and stress fluctuations
have been recorded in microplasticty \cite{uchdim,lee07}, 
it is not clear if they have the same
origin as in fracture. Here, we measure the yield stress
distribution for the PP and ME models. Figure~\ref{fig2} shows the rescaled
yield stress
cumulative distributions for both ME and PP problems with free and pinned
boundary conditions. The latter corresponds to the usual ME problem studied in
the literature while the ``free'' case is closer to experimental reality. We see
that for both boundary conditions the distributions for PP and ME 
collapse with the same exponent into a very similar scaling function. 
Since for the ME problem with pinned boundary conditions
 we know that asymptotically the scaling 
function should converge to the Tracy-Widom distribution~\cite{monthus}, we can
speculate
that this is also true for PP. We have also checked that Weibull and other
extremal
distributions are not appropriate to fit the data.

\begin{figure}
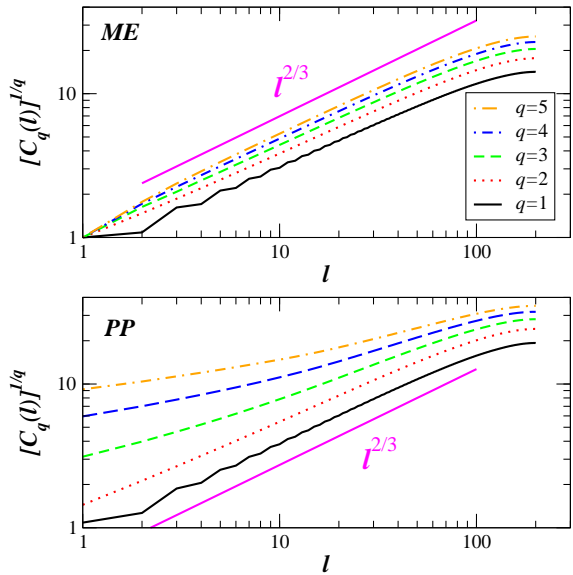

\centerline{\includegraphics *[width=75mm,type=eps,ext=.eps,read=.eps]{fig3}}
\caption{(Color online) Height-height correlation function of order $q$ = 1 to
$q$ = 5 for ME
and PP problems in a system of size $L=200$. Multiscaling of the surface
fluctuations for PP is clearly
observed.}
\label{fig3}
\end{figure}

\begin{figure}
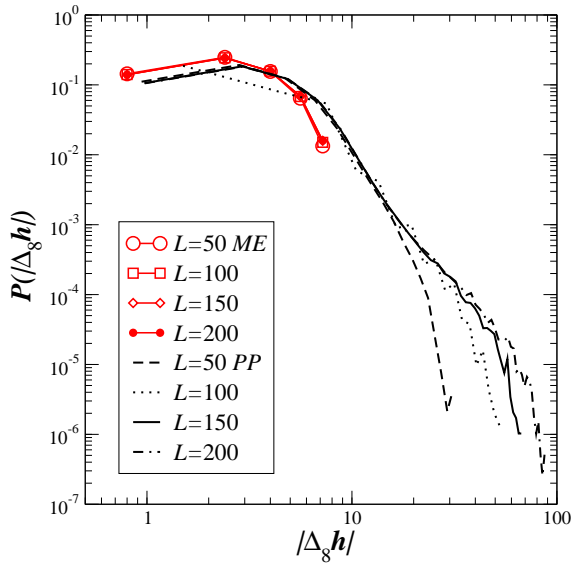

\centerline{\includegraphics *[width=75mm,type=eps,ext=.eps,read=.eps]{fig4}}
\caption{(Color online) Distribution of height differences at a fixed distance
$\ell=8$ in PP
and ME for different system sizes. For PP the tail of the distribution grows
with system size, while it remains constant for ME surfaces.}
\label{fig4}
\end{figure}

The spatial properties of the yield surfaces are analyzed in Fig.~\ref{fig3},
where we
show the $q$th order correlation functions, $C_q(\ell) = \langle
\overline{\vert{h(x+\ell)-h(x)}\vert^{q}}\rangle \sim \ell^{\, q\, \zeta_q}$,
for PP and ME surfaces. A univaluated height is constructed by taking the
maximum surface value $h(x)$ at each site $x$. ME surfaces exhibit the expected
"simple" self-affine scaling, that is, the correlation function scales with the
same roughness exponent $\zeta_{q}^{ME}= \zeta^{ME} = 2/3$ for all $q$. This is
in agreement with previous studies showing that overhangs are irrelevant in ME
surfaces below the strong disorder limit~\cite{hav98,hav06}.
In contrast, PP surfaces show strong deviations
from simple self-affinity and the existence of multiscaling becomes readily
evident in Fig.~\ref{fig3} (bottom panel). This indicates that overhangs are
indeed
relevant in PP surfaces. This is illustrated by studying the distribution of
height differences at different length scales $\mathcal{P}\left(|\Delta_\ell
h|\right)$ with $\Delta_\ell h \equiv h\left(x+\ell\right) - h\left(x\right)$.
For a self-affine interface, this distribution is expected to scale as
$\mathcal{P}\left(|\Delta_\ell h|\right)\sim \ell^{-\alpha}f\left(|\Delta_\ell
h| /\ell^{-\alpha} \right)$. To obtain further insight on the role of overhangs
at different scales we analyze the distribution for intermediate values of $\ell
\ll L$. In Fig.~\ref{fig4} it is shown that for ME surfaces
$\mathcal{P}\left(|\Delta_8 h|\right)$ is narrow and independent of $L$, whereas
for the PP surfaces the tail grows with $L$ and approaches asymptotically a
power-law
shape, $\mathcal{P}\left(|\Delta_8 h|\right) \sim |\Delta_8 h|^{-2}$.

In summary, we have shown that the principle of load-sharing in a yielding
material introduces the ``yield surfaces'' as a separate statistical mechanics
problem. Our main result is that ME and PP correspond to two different
optimization problems in disordered media. The reason for the non-equivalence
between ME and PP surfaces arises from the fact that an actual yield surface---
with signed currents--- is created in a yielding material before the ME surface.
This is intimately related to the peculiar properties of PP surfaces such as
relevant overhangs, large hight-height fluctuations, and lack of simple
self-affinity. In addition, the yield stress displays a finite-size scaling form
with corrections due to the boundary conditions.  It would be interesting to
study numerically more realistic models of plasticity, and to investigate the
role of dimensionality since in three dimensions the large surface fluctuations
are theoretically expected to diminish.

\acknowledgments
CBP and JML acknowledge financial support from the Ministerio de Ciencia e
Innovaci{\'o}n (Spain) under project FIS2009-12964-C05-05. 
CBP is supported by
a
FPU fellowship (Ministerio de Ciencia e Innovaci{\'o}n, Spain) and thanks the
ISI foundation for hospitality and support. MJA would like to acknowledge the
support of the Center of Excellence program of the Academy of Finland. MJA and
SZ gratefully thank the financial support of the European Commission NEST
Pathfinder programme TRIGS under contract NEST-2005-PATH-COM-043386.


\end{document}